\documentclass[preprint,prd,aps,showpacs,preprintnumbers,amssymb,amsmath]{revtex4}

\usepackage{natbib}
\usepackage{mathrsfs}
\usepackage{graphicx}

\usepackage{latexsym}
\usepackage{epsfig}
\usepackage{amscd}
\usepackage{psfrag}
\usepackage{tabularx}

\begin{document}

\title{Numerical Simulations of a Spherically Symmetric Yang-Mills system on a de Sitter Manifold}

\author{H. Lux}
\affiliation{Department of Theoretical Physics, University of Z\"urich, Winterthurerstr. 190, CH-8057 Z\"urich, Switzerland}
\email{lux@physik.unizh.ch}

\author{K. Johannsen}
\affiliation{Institute for Computational and Applied Mathematics, University of Bergen, Thorm\o hlensgate 55, N-5008 Bergen, Norway}
\email{klaus.johannsen@bccs.uib.no}

\date{\today}

\begin{abstract}
In this paper we discuss the dynamics of the cosmological Bartnick-McKinnon analogue with $n=1$ and $\Lambda=\Lambda_{reg}(n)$ . We derive boundary conditions from energy considerations. Numerical simulations are carried out to show the existence of a non-stationary solution.
\end{abstract}

\pacs{02.70.Dh, 04.62.+v, 11.27.+d}

%\keywords{Numerical simulations, Yang-Mills, de Sitter, boundary conditions,  classical solutions, FEM}

\maketitle

%%%%%%%%%%%%%%%%%%%%%%%%%%%%%%%%%%%%%%%%%%%%%%%%%%
\section{Introduction}

%Allgemein
In general numerical simulations of the dynamics of soliton solutions in particle physics are of fundamental interest to test their stability. Specifically the dynamics of Einstein-Yang-Mills systems help to understand the interplay of gravity and non-abelian gauge fields. As analytical solutions for the fully nonlinear equations are usually not available in these cases, only numerical simulations can provide further insight.

Bartnik and McKinnon \cite{Bart} found that particle-like solutions for the Einstein-Yang-Mills system exist, while that is not the case for either pure Yang-Mills or the pure Einstein systems \cite{Steve}. Volkov et al. \cite{Volk} investigated the Einstein-Yang-Mills equations with cosmological constant $\Lambda$ to find an analogous class of solutions.

In \cite{Brod} Brodbeck et al. performed the respective stability analysis of these static Bartnick-McKinnon analogues.  First steps in simulating the dynamical evolution of the the ground state $n=1$ in the case of $\Lambda=\Lambda_{reg}(n)$ has been taken in \cite{Ralf}.  Different to the approach of \cite{Ralf} we use a Finite Element scheme for our numerical simulations and a reformulated differential equation. Special emphasis is on the boundary conditions in relation of the conservation of energy. In our approach we consider the static de Sitter manifold only.

The paper is organized as follows. In section~\ref{sec:eq} we discuss the physical field equation that governs the dynamics of the monopole. In
section~\ref{sec:bnd} we derive the initial- and boundary conditions in relation to the particle energy. The discretisation and related stability
issues are discussed in section~\ref{sec:disc}. Finally numerical results are given in section~\ref{sec:results}.

%%%%%%%%%%%%%%%%%%%%%%%%%%%%%%%%%%%%%%%%%%%%%%%%%%%
\section{The physical field equations}
\label{sec:eq}
In their analysis of Einstein-Yang-Mills fields with a cosmological constant Volkov et al. \cite{Volk}  found that three different classes of solutions exist. These are static, spherically symmetric and are characterized by their value of $\Lambda$ and the number of nodes $n$. The three different classes can be distinguished by their values of $\Lambda$. For $\Lambda<\Lambda_{crit}(n)$ the solutions approach asymptotically the de Sitter geometry. The solutions with $\Lambda_{crit}(n) <\Lambda <\Lambda_{reg}(n)$ are the so called "bag of gold solutions". They have finite size, but still an horizon. All the solutions with non-vanishing $\Lambda < \Lambda_{reg}$ have non-vanishing magnetic charge. At $\Lambda_{reg}$ the solutions are compact solutions on a de Sitter manifold, where magnetic charges are no longer defined. 

The smooth, spherically symmetric Yang-Mills field has a SU(2) gauge group. The components of the Gauge field in spherical coordinates for the Abelian gauge are
\begin{subequations}
\begin{alignat}{4}
	&A_{t}&=&0, &\qquad& A_{\vartheta}&=&-\frac{1}{g} T_{2} w(r,t),\\
	&A_{r}&=&0, &\qquad& A_{\varphi}&=&-\frac{1}{g}\left(-T_{1} w(r,t)\sin\vartheta+T_{3}\cos\vartheta\right), 
\end{alignat}
\label{feq}
\end{subequations}
where $g$ is the coupling constant of the Yang-Mills (YM) field and the $T_{i}=\frac{1}{2} \sigma_{i}$ are the generators of the SU(2)
group, $\sigma_{i}$ being the Pauli matrices. This gauge has been chosen as the Yang-Mills field in this case only depends on $r$ and
$t$, which reflects the spherical symmetry of the field. %This gauge can be derived from the Coulomb gauge as given in \cite{Huang} by $U(\vartheta,\varphi)=\exp(i\vartheta T_{2})\exp(i\varphi T_{3})$ and slight modifications. This is analogous to the approach in \cite{GuFo}.

The de Sitter manifold is defined by its metric. In conformal coordinates \cite{Birr, Houch} it is given by
\begin{alignat}{2}
	& \textnormal{diag}(g_{\mu\nu})=\left(\frac{1}{\sin^{2} t}, -\frac{1}{\sin^{2} t},-\frac{\sin^{2} r}{\sin^{2} t},-\frac{\sin^2 r\sin^2 \vartheta}{\sin^{2} t}\right).
\end{alignat}
The coordinates ranges are $t\in ]0,\pi[$, $r\in[0,\pi[$, $\vartheta\in[0,\pi[$ and $\varphi \in[0,2\pi[$.
Its curvature is given by $\Lambda/3$, where $\Lambda$ is the cosmological constant.  The field equations can be derived through using the
variational principle on the appropriate Lagrangian. The Yang-Mills Lagrangian is given by 
\begin{alignat}{2}
	&\mathscr{L}_{YM}&=&-Tr(F_{\mu\nu}F^{\mu\nu}),\qquad\text{with}\\
	&F_{\mu\nu}&=&\partial_{\mu}A_{\nu}-\partial_{\nu}A_{\mu}+ig[A_{\mu},A_{\nu}].
\end{alignat}
$F_{\mu\nu}$ is the associated 2-form field of the gauge field and is also called the field strength tensor and $[\cdot,\cdot]$ is the product of
the $\mathfrak{su}(2)$ Lie algebra. The equations for the coupled Einstein-Yang-Mills system have been derived in \cite{Ralf}, as well as
the equations for the decoupled Yang-Mills field on a static de Sitter manifold
\begin{alignat}{2}
	&\ddot{w}-w''+\frac{w(w^{2}-1)}{\sin^2 r}=0.  \label{fieldeq}
\end{alignat}
Here the dot denotes the derivative with respect to time and prime the derivative with respect to $r$. This is a semi-linear partial differential equation of hyperbolic type.

%%%%%%%%%%%%%%%%%%%%%%%%%%%%%%%%%%%%%%%%%%%%%%%%%
\section{Initial and boundary conditions}
\label{sec:bnd}

In this paper we investigate the numerical solution of equation \eqref{fieldeq}. It is defined on the compact domain 
\begin{alignat}{2}
    &\Omega=I\times I, \qquad I=[0,\pi]_{(\mathbb{R}}, \notag
\end{alignat}
and requires initial conditions for both $w$ and $\dot w$. For the initial conditions for $w$ we will chose the stationary solution given in equation \eqref{eqinit},
for the initial conditions for $\dot w$ we introduce a perturbation of the stationary system. Boundary conditions follow from energy
considerations, see below. 

\subsection{Initial Conditions}
As we have already mentioned before we are analysing the dynamical evolution of the static Bartnik-McKinnon analogue with $n=1$ and $\Lambda=\Lambda_{reg}=3/2$. It is given by
\begin{alignat}{2}
	&w_0=cos(r),\quad r\in I.  \label{eqinit}
\end{alignat}
Thus it represents our initial condition for $w$. In the following we set the coupling constant
$g=1$. As the initial condition for~$\dot w$ we choose
\begin{alignat}{1}
	&\dot{w}_0=\sin(r)\dot f_0(r,t)\in L_2(I), \label{teqinit}
\end{alignat}
with an arbitrary function $f_0$, which was chosen in different numerical experiments differently.

\subsection{Boundary Conditions}
The boundary conditions for the system will be derived from the finiteness of the energy.
For systems on a Minkowski space-times, these types of boundary conditions are related to the asymptotic behavior of the field 
at infinity. These conditions are implemented by an appropriate choice of the functions-space, e.g. $L_2(\omega)$, etc.
As for the Minkowski space-times, for the compact de Sitter manifold the energy finiteness of $w$ will be related to an appropriate
choice of the function space.
\subsubsection{Energy conditions for $\vartheta$, $\varphi$.}  As we are investigating a spherical symmetric Yang-Mills field the energy is independent of these variables.
\subsubsection{Energy Conditions for $r$.}  In \cite{Chru} Chrusciel et al. derived a priori estimates of the solution space of Yang-Mills equations on globally hyperbolic four dimensional Lorentz manifolds. They show that the Yang-Mills gauge field $A_{\mu}$ belongs to the Sobolev space $H^{2}(\mathbb{R}^3)$ with
\begin{alignat}{1}
    & \|A_{\mu}(\cdot,t)\|_{H^{2}(\mathbb{R}^3)}\leq C(t)\notag
\end{alignat}
and an appropriate constant $C(t)\leq\infty$. From the assumed symmetry ($A_{\mu}$ is independent of $\vartheta$, $\varphi$) and \eqref{feq}
it follows $w(\cdot,t)\in H^{1}(I)\cap\mathbb{C}^{0,1/2}(I)$ with
\begin{alignat}{1}
    & \|w(\cdot,t)\|_{H^{1}(\mathbb{R})}\leq C(t).\label{wreg}
\end{alignat}
The energy of the Yang-Mills field in the decoupled case is given by
\begin{alignat}{1}
	E(t)=\int_{t=const.} T_{\mu\nu} n^{\mu}\xi^{\nu} dV,\notag
\end{alignat}
where $T_{\mu\nu} = F_{\mu\lambda}{F^{\lambda}}_{\nu}-\frac{1}{4}g_{\mu\nu}F_{\lambda\sigma}F^{\lambda\sigma}$is the energy momentum tensor
and $n^{\nu}=(\sin t ,0,0,0)$, $\xi^{\nu}=(1,0,0,0)$ are the normal of the domain pointing outside and a Killing vector respectively. With
the constant of proportionality $C$ we rewrite the energy as
\begin{alignat}{2}
	&E(t)&=&C\int_0^\pi \left( w'^2(r,t)+\dot{w}^2(r,t)+\frac{(1-w^2(r,t))^2}{2\sin^2 r}\right) dr\notag \\
    &&=&C\left(\|w(\cdot,t)\|_{H^1}^2+\|\dot{w}(\cdot,t)\|_{L_2}^2+\frac{1}{2}\left\|\frac{1-w^2(\cdot,t)}{\sin(\cdot)}\right\|_{L_2}^2\right).\label{energy}
\end{alignat}
Due to the continuity of $w(r,t)$ with respect to $r$, the finiteness of $E(t)$ implies $w(r,t)=\pm1$, $\forall r=0,\pi$ and $t\in I$.
We choose
\begin{alignat}{1}
    &w(0,t)=1, \, w(\pi,t)=-1, \qquad\forall t\in I.\label{bc}
\end{alignat}
Assuming the boundary conditions  \eqref{bc} and a field $w(r,t)$ governed by equation \eqref{fieldeq}, it is easily varified that the energy is conserved at all times. With \eqref{eqinit} it follows
\begin{alignat}{2}
    &E(t)=E(0)=C\left(3\pi/4+\|\dot{w}_0\|_{L_2}^2\right).\label{energy2}
\end{alignat}
As for the initial condition \eqref{eqinit}, we assume vanishing spatial derivatives of the solution at $r=0,\pi$ for all times.
This follows from the assumption that the solution should describe a smooth solution on the three-dimensional sphere requiring
tangential planes at the poles corresponding to $r=0,\pi$. We therefore make the ansatz
\begin{alignat}{1}
    w(r,t)=\cos(r)+\sin(r)f(r,t),\quad (r,t)\in\Omega \label{ansatz}
\end{alignat}
with the unknown function $f(r,t)$. From \eqref{fieldeq} follows the equation for $f$
\begin{eqnarray}
\nonumber
	\sin(r)(\ddot{f}-f'')-2\cos(r)f'+2\frac{\cos^2(r)}{\sin(r)}f&&\\
	+3\cos(r)f^2+\sin(r)f^3&=&0,\label{eq4f}
\end{eqnarray}
with the boundary and initial conditions
\begin{alignat}{2}
\nonumber
    &f(0,t)=f(\pi,t)=0,\,\forall t\in I, \\
    &f(r,0)=0,\,\dot f(r,0)=\dot f_0(r),\,\forall t\in I. \label{init4f}
\end{alignat}
The following two terms in equation \eqref{eq4f} need special attention: The scaling factor of d'Alem\-bert operator ($\sin(r)$ vanishes
at the boundary $r=0,\pi$ and the source term linear in $f$ has infinite weights at the boundaries. The first difficulty
is in view of a non-vanishing source at the boundary unproblematic. Assuming the same regularity for $f$ as we have for $w$,
i.e. $f(\cdot,t)\in H^{1}(I)\cap\mathbb{C}^{0,1/2}(I)$, the linearly diverging weight of the linear source term is also not
problematic.

%%%%%%%%%%%%%%%%%%%%%%%%%%%%%%%%%%%%%%%%%%%%%%%%%%%%%%%
\section{Discretisation}
\label{sec:disc}

To integrate \eqref{eq4f} forwards in time, we transformed the equation to a set of first order equations, which we integrate by two different implicit
schemes. The first is the classical implicit Euler methods, the second a modification of a two-stage Runge-Kutta method (Alexander scheme)\cite{Klaus}, first
introduced by in \cite{Alex}.

For the discretisation of the spatial dimension we chose the conforming Finite Element discretisation with continuous, piece-wise linear
basis functions.  We used a equidistant grid with mesh-size $h=\pi/n_r$, $n_r>0$. The spatial integrals have been carried out with \emph{Maple}
with the exception of the fourth term in equation \eqref{eq4f}. This term cannot be integrated analytically and has been treated by appropriate integration rules (Simpson rule). The numerical approximation of these integrals has been investigated and chosen appropriately to avoid numerical artifacts.

Additionally it was necessary to use Taylor expansions for the sine and cosine terms in the discretisation of the two non-linear terms in f. As an illustration for this issue we discuss the term cubic in f. Its discretisation is given by
\begin{alignat}{2}
	(\sin r\ (f_{i}\psi_{i})^{3},\psi_{j}),\notag
\end{alignat}
where $(\cdot,\cdot)$ denotes the $L_{2}$ scalar product, $\psi_{i}$ and $\psi_{j}$ denote the piecewise linear basis functions. 
For $j=0$ it has the form $C_1f_0^3+C_2f_0^2f_1+C_3f_0f_1^2+C_4f_1^3$. All the terms are similar, so we concentrate on
\begin{eqnarray}
\nonumber
	C_4&=&\frac{1}{h^4}[-24\cos(h) +6 h \sin(h)+24 \cos(2h) -h^3\sin(2h) \\
	\nonumber
	&&-6h^2\cos(2h)+18h\sin(2h)].
\end{eqnarray}
Here special attention is required to control round-off errors. To accurately account for the difference $24 \cos(2h)-24\cos(h)$ an expansion
of the trigonometric functions had to be carried out. If was found that an expansion up to the sixth order in $h$ was sufficient, leading to
an evaluation of $C_4$ according to
\begin{alignat}{2}
	C_4=\frac{h^2}{12}-\frac{67h^4}{1680}+\frac{257h^6}{43200}+O(h^8).\notag
\end{alignat}
Numerical experiments have shown that a further increase of the order were not necessary.

%%%%%%%%%%%%%%%%%%%%%%%%%%%%%%%%%%%%%%%%%%%%%%%%%%%%%%%
\section{Numerical Results and Conclusions}
\label{sec:results}

The simulations have been carried out with a software tool \emph{sEYMs-1D} (\underline{s}imple \underline{E}instein-\underline{Y}ang-\underline{M}ills 
\underline{s}olver for one dimension), a C++-code. A documentation is available with \cite{Hanni}.

\begin{figure*}[h]
	\centering
	\includegraphics[width=\textwidth]{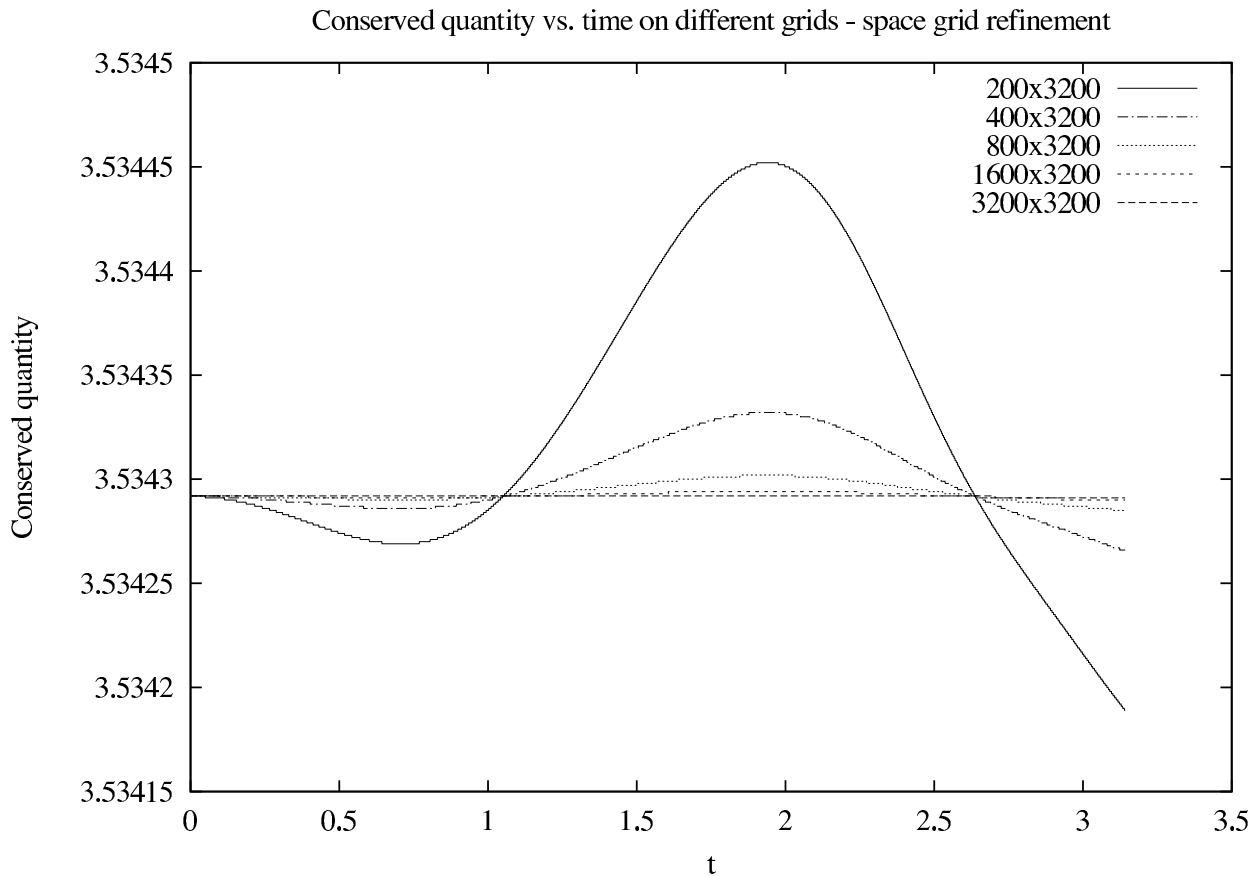}
	\caption{Energy conservation of the YM field. The field $f(r,t)$ has been computed using the Alexander scheme for time integration with
	initial conditions $\dot f_0=\sin(r)$. The label of the graphs indicate the spatial and temporal resolution $n_r\times n_t$, where
	$n_r$ and $n_t$ denote number of elements in the spatial and temporal direction, respectively.}
	\label{fig:EnHV}
\end{figure*}

Several numerical experiments have been carried out. To validate to code, the conservation of energy has been considered. Its behavior over
time is displayed in Fig.~\ref{fig:EnHV} on a sequence of successively refined grids.

\begin{figure*}[h]
\centering
\includegraphics[width=\textwidth]{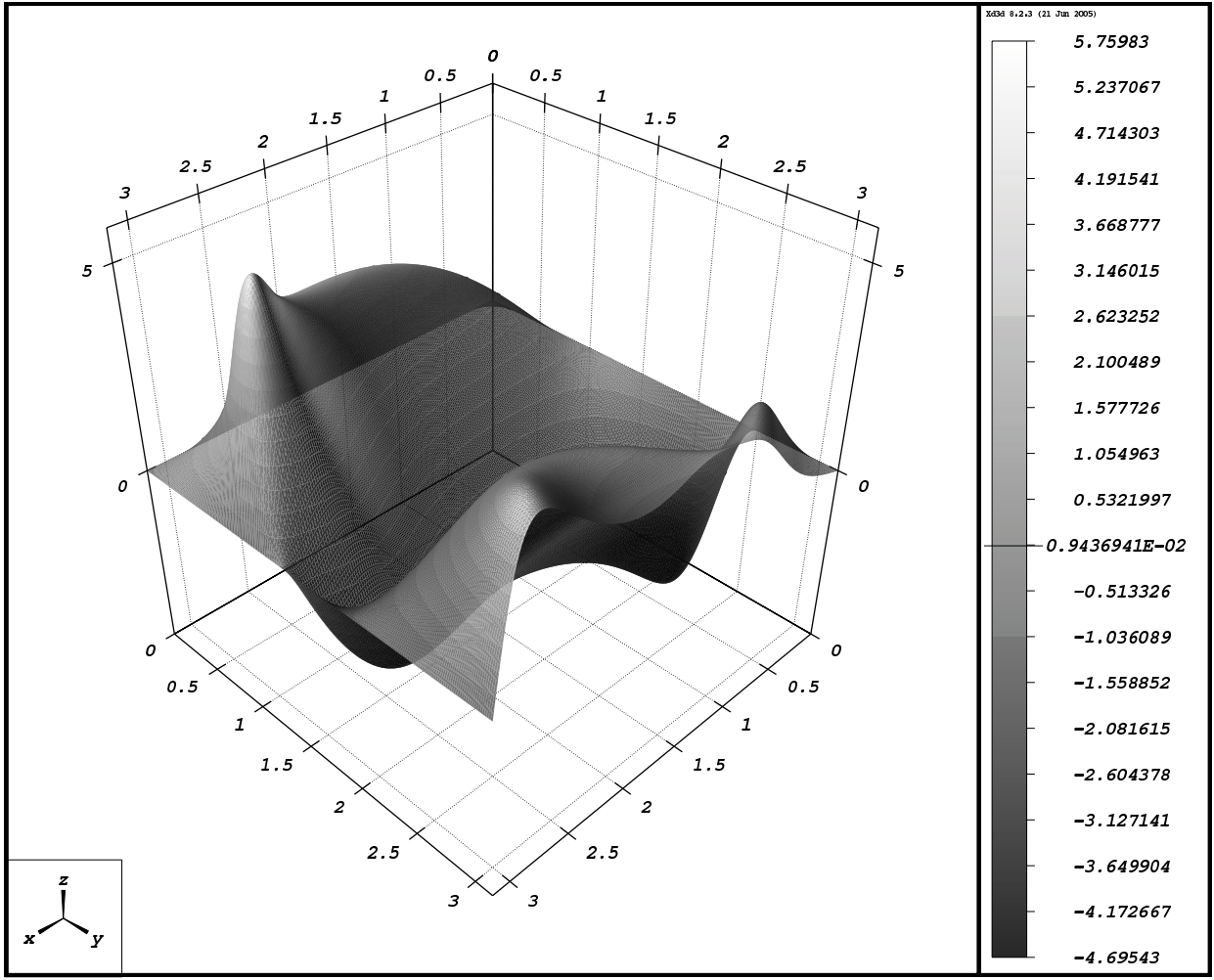}
\caption{$f(r,t)$ for initial conditions $\dot f_0=10\sin(r)$. $r$, $t$ and $f(r,t)$ are varying along the x-, y- and z-axis, respectively}
\label{fig:s10}
\end{figure*}

\begin{figure*}[h]
\centering
\includegraphics[width=\textwidth]{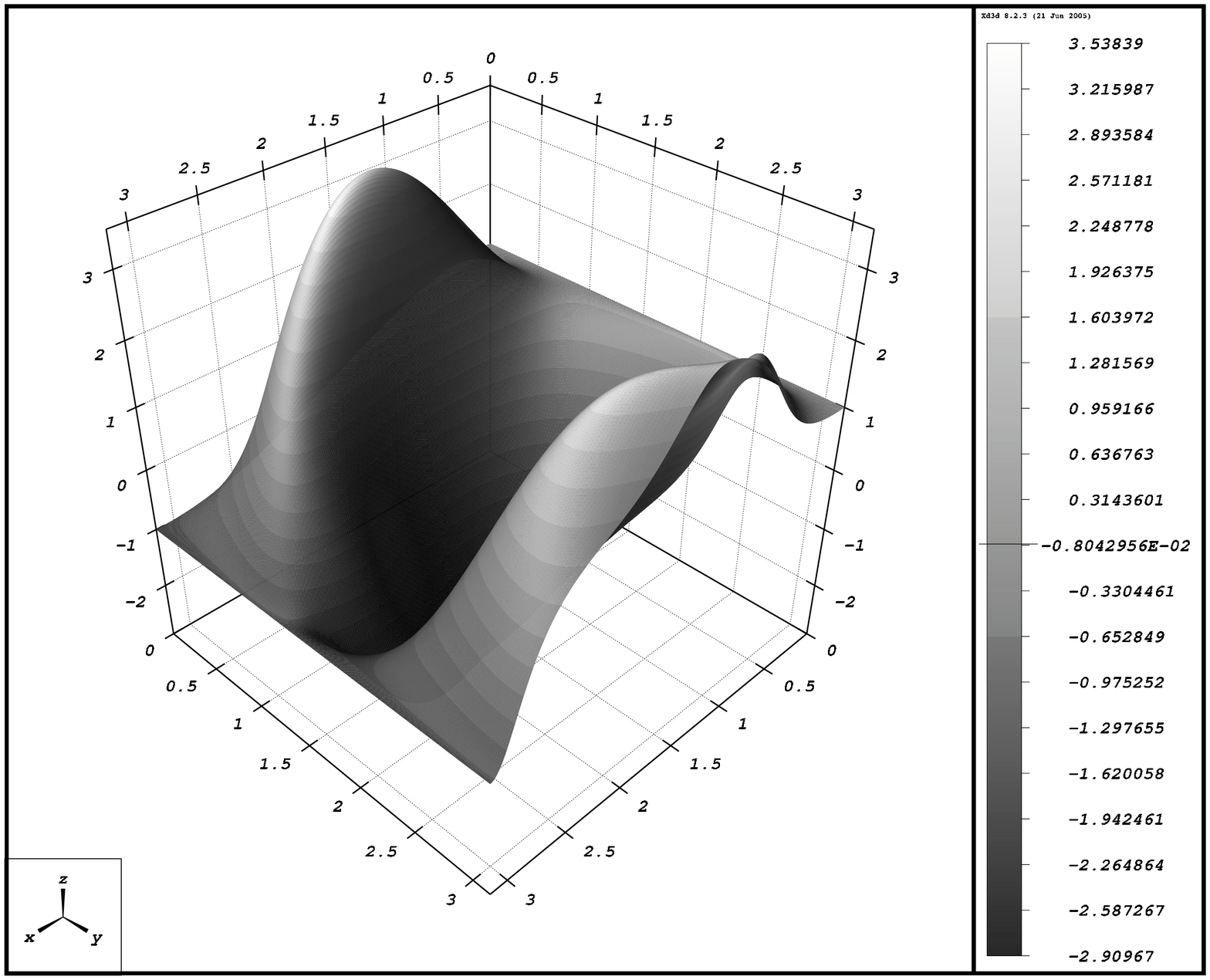}
\caption{$w(r,t)$ for initial conditions $\dot f_0=10\sin(r)$. $r$, $t$ and $w(r,t)$ are varying along the x-, y- and z-axis, respectively}
\label{fig:s10w}
\end{figure*}

As a second numerical example we investigated the behavior of the field resulting from the initial conditions \eqref{init4f} with
\begin{alignat}{1}
	& \dot f_0=10\sin(r).
\end{alignat}
The amplitude has been chosen large enough in order to show the full non-linear dynamics of the system without destroying the original 
structure of the field. The simulations has been carried out on a $200\times 200$ grid (using same notation as above) using the 
Implicit Euler method. It is displayed in Fig.~\ref{fig:s10}. The result in terms of the original field $w$ is shown in Fig.~\ref{fig:s10w}.
The visualization has been carried out with the program \emph{xd3d}

As can be seen clearly, energy transfer takes place between the different modes resulting in a strongly perturbed steady-state solution.
No plane wave solutions can be observed. This is a property intrinsic to Yang-Mills fields and has already been seen on flat manifolds.

%%%%%%%%%%%%%%%%%%%%%%%%%%%%%%%%%%%%%%%%%%%%%%%%%%%%%%%
\begin{acknowledgments}
Firstly the authors would like to thank the referee for his valuable comments. Furthermore we would like to thank Matthias Bartelmann, J\"org Frauendiener and Ralf Peter for continuous support and helpful discussions. Additionally we would like to thank Jonathan Coles for finding the "last" programming error and Thomas Gehrmann for his helpful remarks.
\end{acknowledgments}

\end{document}